\documentclass[12p]{elsarticle}
\usepackage{lineno,hyperref}
\modulolinenumbers[5]
\journal{Physica A: Statistical Mechanics and its Applications}
\begin{document}
\begin{frontmatter}
\title{Quantification of scaling exponent with crossover type phenomena for different types of forcing in DC glow discharge plasma}
\author{Debajyoti Saha \footnote{Corresponding author. Tel : +91 9775429915; E-mail address : debajyoti.saha@saha.ac.in}}
\author{Pankaj Kumar Shaw}
\author{Sabuj Ghosh}
\author{M. S. Janaki}
\author{A. N. Sekar Iyengar}
\address{Plasma Physics Division, Saha Institute of Nuclear Physics, HBNI, 1/AF, Bidhannagar, Kolkata - 700064, India.}

\begin{abstract}
We have carried out a detailed study of scaling region using detrended fractal analysis test by applying different forcing likewise noise, sinusoidal, square on the floating potential fluctuations acquired under different pressures in a DC glow discharge plasma. The transition in the dynamics is observed through recurrence plot techniques which is an efficient method to observe the critical regime transitions in dynamics. The complexity of the nonlinear fluctuation has been revealed with the help of recurrence quantification analysis which is a suitable tool for investigating recurrence, an ubiquitous feature providing a deep insight into the dynamics of real dynamical system. An informal test for stationarity which checks for the compatibility of nonlinear approximations to the dynamics made in different segments in a time series has been proposed. In case of sinusoidal, noise, square forcing applied on fluctuation acquired at P=0.12mbar only one dominant scaling region is observed whereas the forcing applied on fluctuation (P=0.004mbar) two prominent scaling regions have been explored reliably using different forcing amplitudes indicating the signature of crossover phenomena. Furthermore a persistence long range behaviour has been observed in one of these scaling regions. A comprehensive study of the quantification of scaling exponents has been carried out with the increase in amplitude and frequency of sinusoidal, square type of forcings. The scalings exponent is envisaged to be the roughness of the time series. The method provides a single quantitative idea of the scaling exponent to quantify the correlation properties of a signal.
\end{abstract}
\end{frontmatter}

\section{Introduction}

In recent years there has been growing evidence that many physical and biological system  have no characteristic length scale and exhibit a power law correlation. Traditional approaches such as power spectrum, correlation analysis are suited to quantify correlations in a stationary signal \cite{stationary,stationary1}. However many signals that are output of complex physical \cite{Haus} and biological system are said to contain nonstationarity. Almost all methods of time series analysis, traditional linear or nonlinear, must assume some kind of stationarity \cite{epl}.
A number of statistical tests for stationarity \cite{peng,peng1} in a time series have been proposed in the literature.
Most of the tests we are aware of are based on ideas similar to the following: Estimate a certain parameter using different parts of the sequence.
If the observed variations are found to be significant, that is, outside the expected statistical fluctuations, the time series is regarded as nonstationary.
In case of traces of nonstationarity being  detected, we are allowed to carry out modified root mean square analysis termed as detrended fluctuation analysis (DFA) \cite{DFA}. The advantages of DFA is that it permits the detection of long range correlation embedded in a seemingly non- stationary time series and allows the detection of scaling exponent in noisy signal with embedded trend that can mask the true correlations. From a practical point of view if the fluctuations driven by uncorrelated stimuli can be decomposed from the intrinsic fluctuations generated by dynamical system, then these two classes of fluctuations may be shown to have different correlation properties. In the last one decade DFA has emerged as an important technique to study scaling and long range temporal correlation in a nonstationary time series \cite{hamilton} which has been extensively studied in literature. It has been successfully applied to the diverse areas of research such as DNA \cite{DNA, DNA1}, neuron spiking \cite{neuron}, heart rate dynamics \cite{heart,heart1}, economical time series, long time weather report \cite{weather}. DFA is based on the idea that that if the time series contains nonstationarities then the variance of the fluctuations can be studied by successively detrending using linear quadratic, cubic higher order polynomial in a piecewise manner. Most real time series exhibit persistence i.e subsequent element of the time series are correlated \cite{jamanda}. The study of the self similarity and scaling in physics, socio economic sciences in the last several years has brought in new insights and new ideas for modeling them. For instance one of the important empirical results of the market dynamics is that the probablity distribution of price returns r in a typical market displays a power law \cite{DFA1} i.e $P(r)\sim r^{\alpha}$ where $\alpha$ =3. Similar power laws appear for the cumulative frequency distribution of earthquake magnitudes \cite{DFA2}. While the spectral analysis (Fourier method) , wavelet transform modulas maxima (WTMM) analyze the time series directly the DFA is based on the random walk theory, similar to the Hurst rescale range analysis. Presence of strong trends associated with nonstationarity can lead to the false detection of scaling exponent. So for the reliable detection of power law exponent ($\alpha$) it is essential to distinguish trends \cite{DFA3} from the long range fluctuation \cite{kantel,pankaj} intrinsic in the data. Most economic and financial time \cite{eco, eco1,eco2}series are persistent with $\alpha>$0.5. Changes in the dynamics during the measurement period usually constitute an undesired complication of the analysis. In many applications of linear (frequency based) time series analysis \cite{henriabar}, stationarity has to be valid only up to the second moments (“weak stationarity”). Then the obvious approach is to test for changes in second or higher order quantities,
like the mean, the variance, skewness, kurtosis. The purpose of this paper is to discuss the matters
pertaining to the study of the scaling exponent by applying different forcing amplitude along with the reliable detection of the change in the value of scaling exponent in different scaling region characterised as crossover phenomena.

Recurrence plot (RP) and Recurrence quantification analysis \cite{vramori, vramori1}
is a relatively new and advanced technique which helps to identify different dynamical properties of a system under consideration. They have been extensively used in diverse fields such as earth science, plasma, earth science, economy to gain understanding about the nonlinear dynamics of complex system. It has also been utilized as an emerging tool to analyze simulation data of ion temperature gradient turbulence \cite{tur} and dissipative trapped electron mode turbulence \cite{tur1} and to characterize transport dynamics. The knowledge of transition between chaotic, laminar or regular behaviour is essential to understand underlying mechanism behind a complex system \cite{complex}. While linear approaches are not appropriate, there are several nonlinear methods that can suitably account for the transition. The traditional recurrence quantification analysis (RQA) that does not require long time observation is one such appropriate nonlinear method. We can also graphically detect different patterns and structural changes in time series data using recurrence plot (RP's) which exhibit characteristic large and small scale patterns caused by the typical dynamical behaviour \cite{webber}.

The paper is organized as follows. In section II, we present a brief schematic of the experimental setup, followed by the results of the analysis of
Floating potential fluctuation, Recurrence plot, Recurrence quantification analysis in section III. Section IV represent a comprehensive
analysis of nonstationarity with detrended fluctuation analysis (DFA). The results of the analysis for DFA method to estimate the scaling exponent for
different forcing amplitude along with the exploration of crossover phenomena have been carried out in section V. Conclusions are presented in section VI.

\section{Experimental Setup}

\begin{figure}
\centering
\includegraphics[width=9cm,height=6cm]{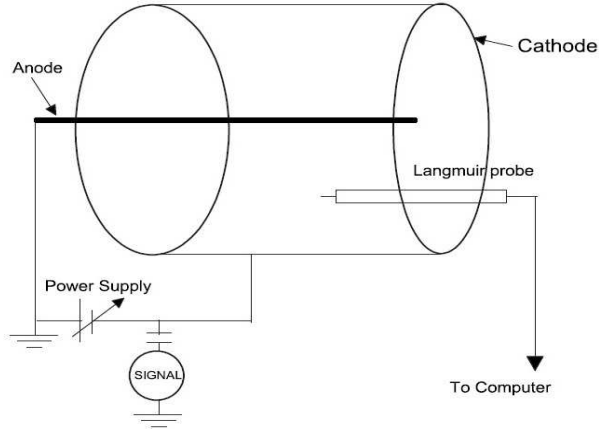}
\caption{Experimental setup for glow discharge plasma}
\label{exp}
\end{figure}

The experiments were carried out in a  cylindrical hollow cathode DC glow discharge argon plasma
with a typical density and temperature of $\sim 10^7/cm^3$ and 2-6
eV respectively. The chamber was evacuated by rotary pump to attain a base pressure of $0.001$ mbar.
Experiments were performed under different forcing amplitudes like sinusoidal, square, noise for two operating neutral pressure i.e 0.12mbar,
0.004mbar respectively with discharge voltage (DV) being fixed at 435Volt. An unbiased Langmuir probe was used to obtain the floating potential
fluctuations acquired with a sampling time of $1*10^{-6}$ sec. A signal generator was also coupled with the DV through a capacitor for observing
fluctuations in presence of forcing as shown in the schematic diagram of Fig. \ref{exp}.

\section{Floating potential fluctuation, Recurrence plot, Recurrence quantification analysis}

\begin{figure}
\centering
\includegraphics[width=14cm,height=8cm]{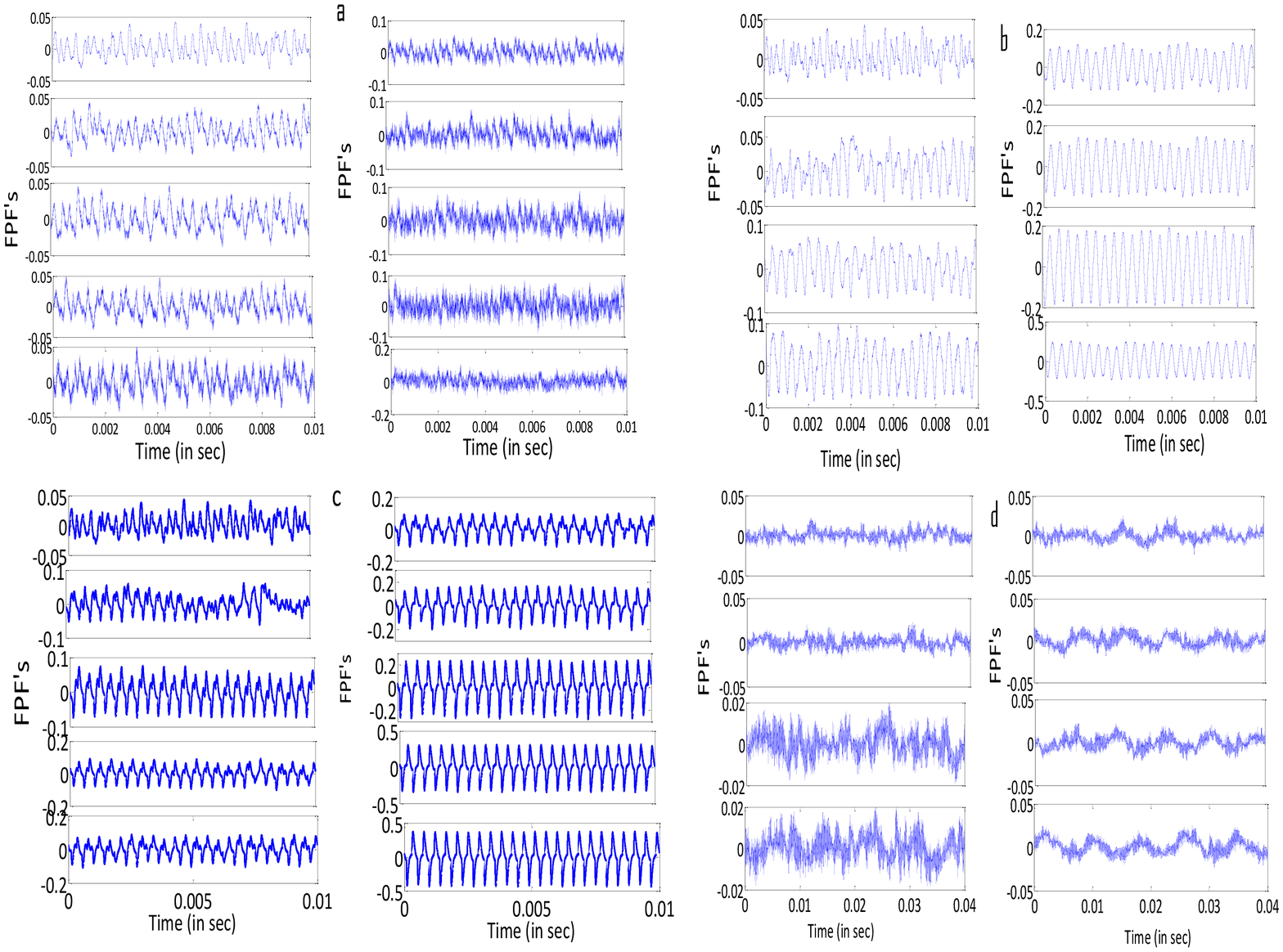}
\caption{Raw signal with increase in amplitude of the a) noise, b) sinusoidal, c) square forcing and d)sinusoidal forcing applied on  fluctuation (P=0.004mbar)}
\label{rr}
\end{figure}

The sequential change in floating potential fluctuations (FPF's) acquired by applying noise, sinusoidal, and square forcing is presented in Fig. \ref{rr}.
The dynamical change in the FPF's has been detected with recurrence plots and its quantification measure namely recurrence quantification analysis (RQA) and discussed in detail with the help of these. We have utilized RP's and RQA's as an important tools for identifying the dynamical transition for different forcing amplitudes. The recurrence plot (RP) \cite{Ecman} is a relatively new technique of time series signals to understand the hidden insights
involving the intricacies of the interplay between nature of different periodicity of the system. The RP expressed as a two dimensional square matrix
represents the occurrence with ones and zeroes for states $X_{i}$ and  $X_{j}$ and find the hidden periodicities in a time series signal which is not
observable by naked eye.

\begin{eqnarray}
R_{ij}=H(\epsilon-||X_i-X_j||)
\label{recur}
\end{eqnarray}

where H is the heavyside function
and $||.||$ is the norm (Euclidean norm), $\epsilon$ is the choice of the threshold with 1 percent of point density in our case.

According to Taken’s embedding theorem  using a time series data $X_{i}$ an embedding can be
made using the vector $Y_{i}={X_{i},X_{i+\tau}...X_{i+(d-1)\tau}}$ which represent the original time series embedded into d dimensional phase space with $\tau$ being the delay.  RP's are graphical, two dimensional representations showing the
instants of time at which a phase space trajectory returns approximately to the same regions of phase space. A recurrence is said to occur whenever a
trajectory visits approximately the same region of phase space indicating $R_{ij}=1$, whereas if the state doesnt recur with itself we are left with $R_{ij}=0$. The cut off distance $\varepsilon$ defines a sphere centered at $x_i$. If $x_i$ falls within the sphere , the state will be close to $x_i$
and thus $R_{i,j}$=1. This $\varepsilon$ can be either constant for all  $x_i$ or they can vary in such  a way that the sphere contains a predefined
number of close states.The diagonal line of length l means the segment of the trajectory is rather close during l time steps to another segment of
the trajectory at different time thus relating these lines to the divergence of the trajectory segments. The average diagonal line length is

\begin{eqnarray}
 L= \frac{\sum_{l=l_{min}}^{N}lP_(l)}{\sum_{l=1}^{N}lP_(l)}
 \label{diagline}
 \end{eqnarray}

\begin{figure}
\centering
\includegraphics[width=12cm,height=8cm]{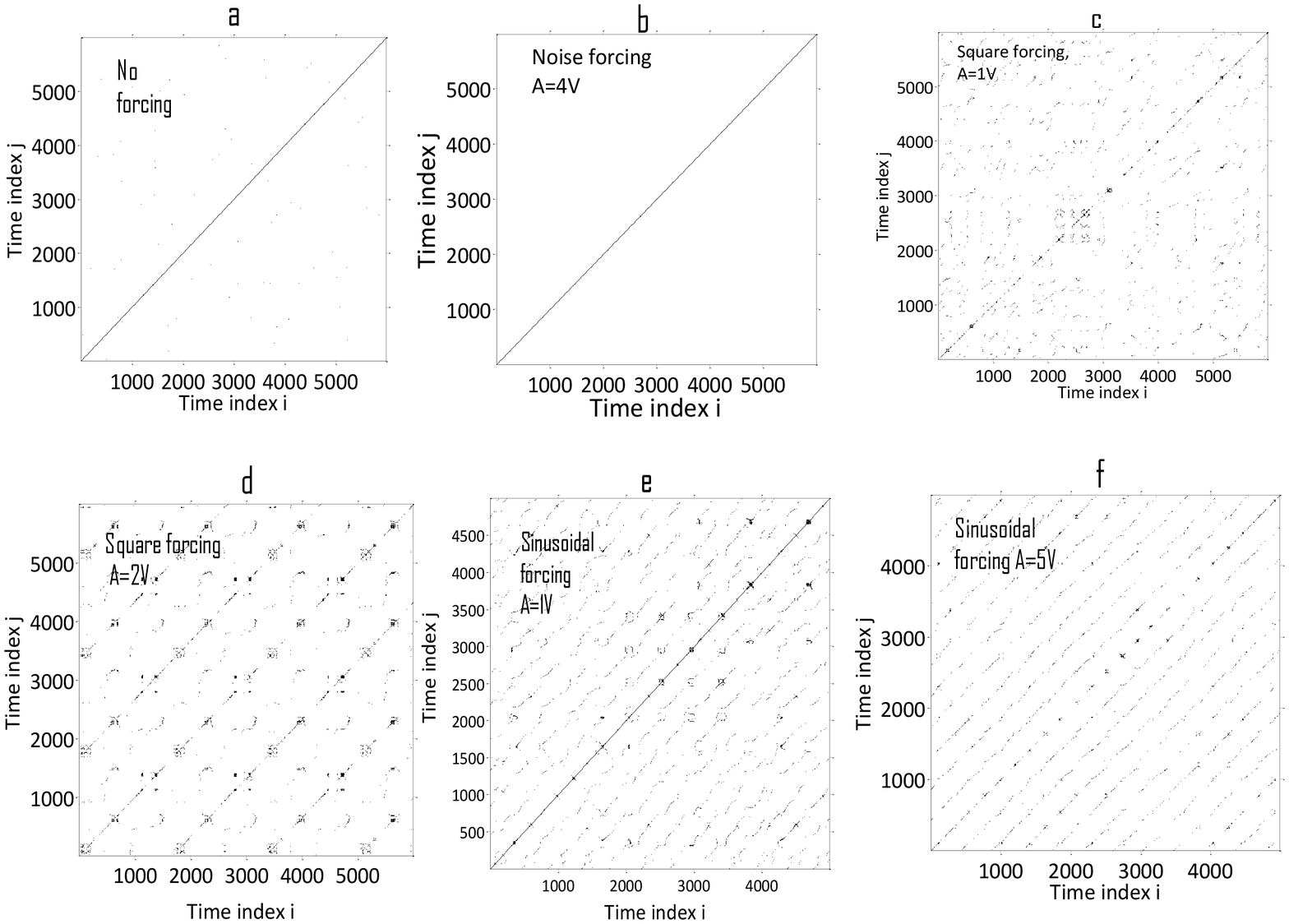}
\caption{Recurrence plots for different types of forcing: a) no forcing b)noise  c)square(A=1V) d)square(A=3V) e)sinusoidal(A=0.8V)
f)sinusoidal A=5Volt}
\label{RP}
\end{figure}

\begin{figure}
\centering
\includegraphics[width=12cm,height=8cm]{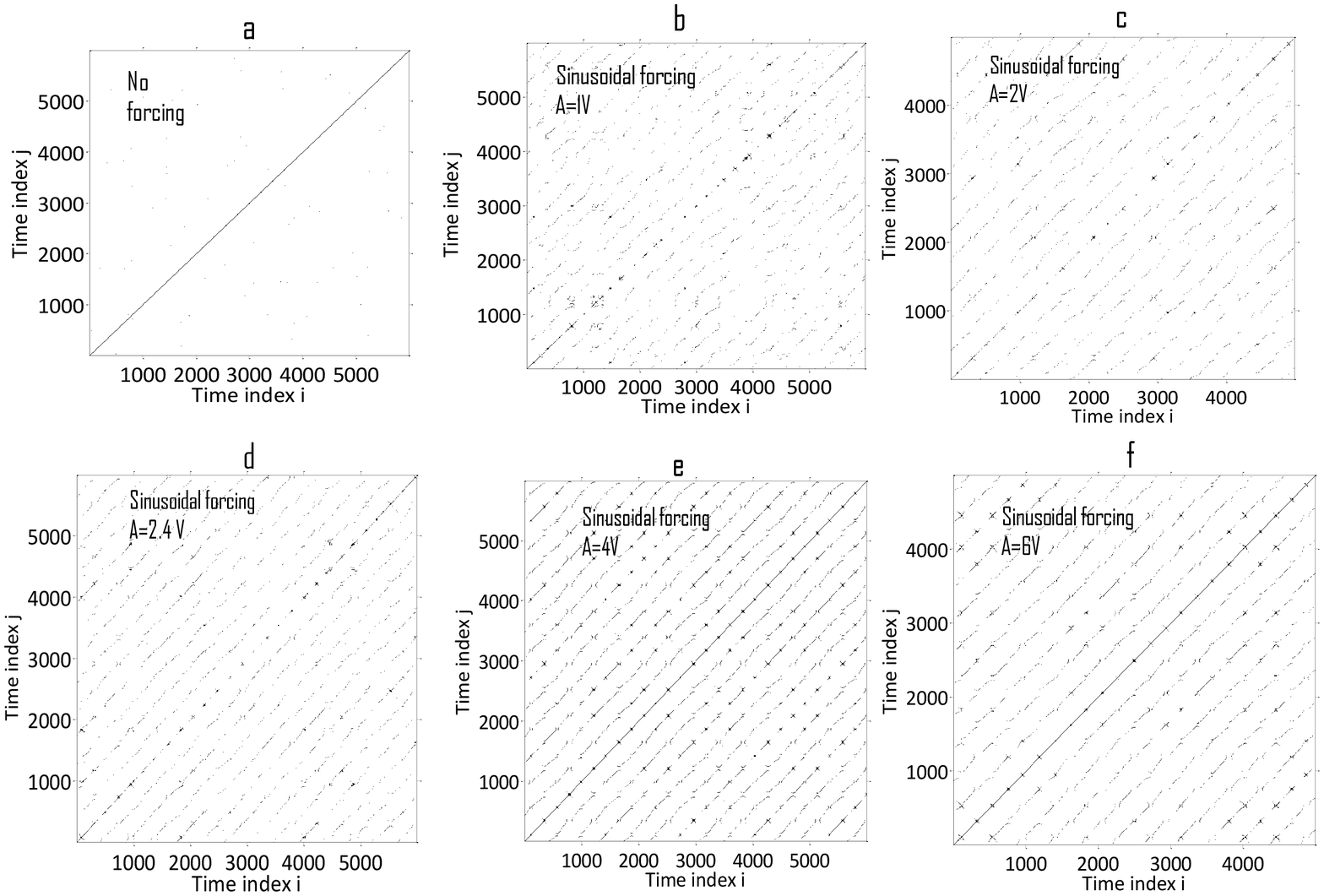}
\caption{Recurrence plot with increase in the amplitudes of sinusoidal forcing (A)}
\label{RP1}
\end{figure}

RP's of FPF's are depicted in Fig.\ref{RP}, Fig \ref{RP1} for different types of external perturbation for the qualitative analysis and visualisation of the
recurrences of dynamical system. Starting with the Fig. \ref{RP}a, \ref{RP}b (corresponding to the time series fluctuation of first and last subplot in the left panel of Fig. \ref{rr}a) without any external forcing and for the noise forcing of amplitude 4V we can hardly observe any point indicating almost zero recurrence followed by the appearance of distinct diagonal lines with scattered points in between long diagonal lines for square forcing of amplitude=1V. The arrangement of the scattered points for square forcing of A=2V (Fig.\ref{RP}d) occupies a large region in between the bold diagonal lines in recurrence plot with increasing number of diagonal lines. The RP in Fig. \ref{RP}e shows a long diagonal line with some faint signature of non-diagonal lines which becomes more ordered and prominent with increasing value of sinusoidal forcing depicted in Fig. \ref{RP}f indicating deterministic behaviour. We have clearly delineated the ordered behaviour in Fig. \ref{RP1} through RP plots in presence of increasing sinusoidal forcing. Initially in Fig. \ref{RP1}b we are left with a main diagonal line with some other faint diagonal lines. The arrangement of the broken diagonal lines along with the scattered points within main long diagonal lines are seen to become more ordered in Fig. \ref{RP1}c, \ref{RP1}d at forcing amplitudes of 2, 2.4V. At higher forcing amplitudes of 5V, 6V RP plots exhibit some prominent arrangement of long diagonal lines.
Diagonal lines in the plots are indicative of periodic, deterministic behaviour and indicate similar evolution of states at different times. Determinism  expressed in the following  equation \ref{dete} is the proportion of recurrence points forming long diagonal structure (of at least length $l_{min}$ to all recurrence points).

\begin{figure}
\centering
\includegraphics[width=12cm,height=8cm]{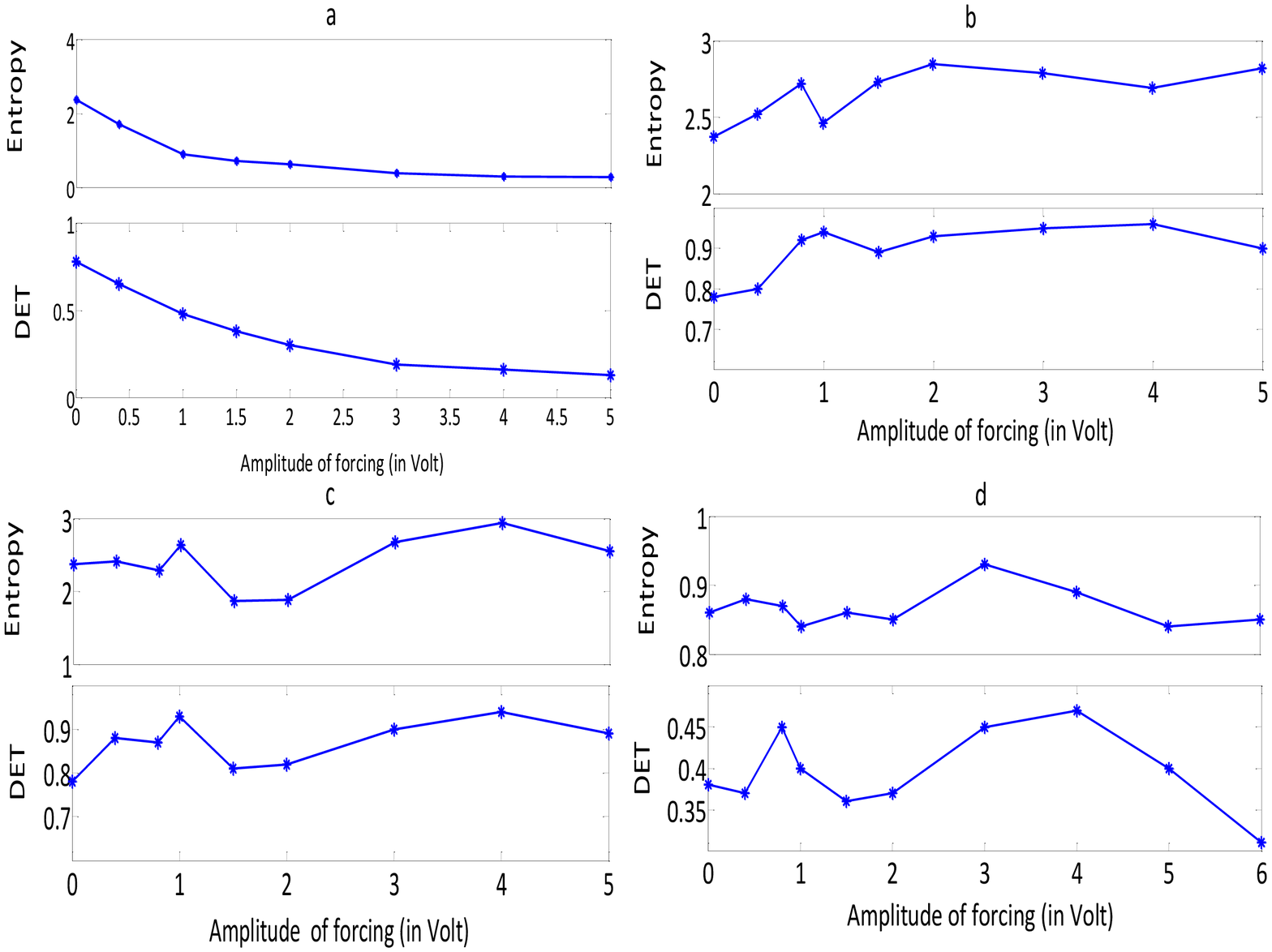}
\caption{Variations of recurrence quantification variables DET, Entropy for increase in a) noise forcing b) sinusoidal forcing c) square forcing d) sinusoidal forcing applied on chaotic fluctuations (P=0.004mbar)}
\label{RQ}
\end{figure}

The recurrence quantification measures determinism (DET), and entropy (ENT) are computed for our experimental
results and plotted with respect to forcing amplitudes of noise, sinusoidal, square forcing respectively in Fig. \ref{RQ}a - \ref{RQ}d. In this work, we are studying the dynamics as well as the statistical property of the plasma fluctuations obtained while applying different types of forcing. Starting with the noise forcing, increase in amplitude of this forcing lead to the constantly decreasing trend in the values of DET, entropy.  DET expressed in following expression \ref{dete} determines whether a signal is periodic or not whereas Entropy illustrates the complexity of the system through the statistics of the diagonal lines lengths in the RP.

\begin{eqnarray}
DET(t)= \frac{\sum_{l=l_{min}}^{N-t}lP_{t}(l)}{\sum_{l=1}^{N-t}lP_{t}(l)}
\label{dete}
\end{eqnarray}

DET is very close to 1 for a purely periodic signal. It has been observed that with increase in the sinusoidal forcing amplitude (Fig. \ref{RQ}b) measures of DET almost shows an increasing trend as the recurrence plots displayed in Fig. \ref{RP1} shows prominent arrangement of uninterrupted diagonal lines increasing DET values. In case of square forcing (Fig. \ref{RQ}c) the DET, entropy values both exhibit same trend showing minima at A=2V corresponding to the RP plot in Fig. \ref{RP}d where the RP plot displays some prominent uninterrupted bold diagonal lines with some scattered points. Here increase in A beyond 2 volt lead to the smooth increasing trend of both DET, Entropy values.  A zigzag pattern in the values of DET for sinusoidal forcing applied on chaotic fluctuation for P=0.004mbar (Fig. \ref{RQ}d) is observed with the occurrence of two local maxima's at A=0.8, 4V. Here the increase in amplitude results in the increasing trend of Entropy upto A=3volt followed by the smooth decreasing trend for the further increase in A. So RQA analysis in corroboration with RP qualitatively and quantitatively reveals the underlying physics of the system dynamics.

\section{Nonstationary, Detrended fluctuation analysis}

Almost all methods of time series analysis, traditional
linear or nonlinear, must assume some kind of stationarity. Therefore, changes in the dynamics during the measurement period usually constitute an undesired complication of the analysis. A number of statistical tests for stationarity in a time
series have been proposed in many literature. Nonlinear statistics which can be used include higher order correlations, dimensions, Lyapunov exponents, and binned probability distributions.. \\

\begin{figure}
\centering
\includegraphics[width=12cm,height=8cm]{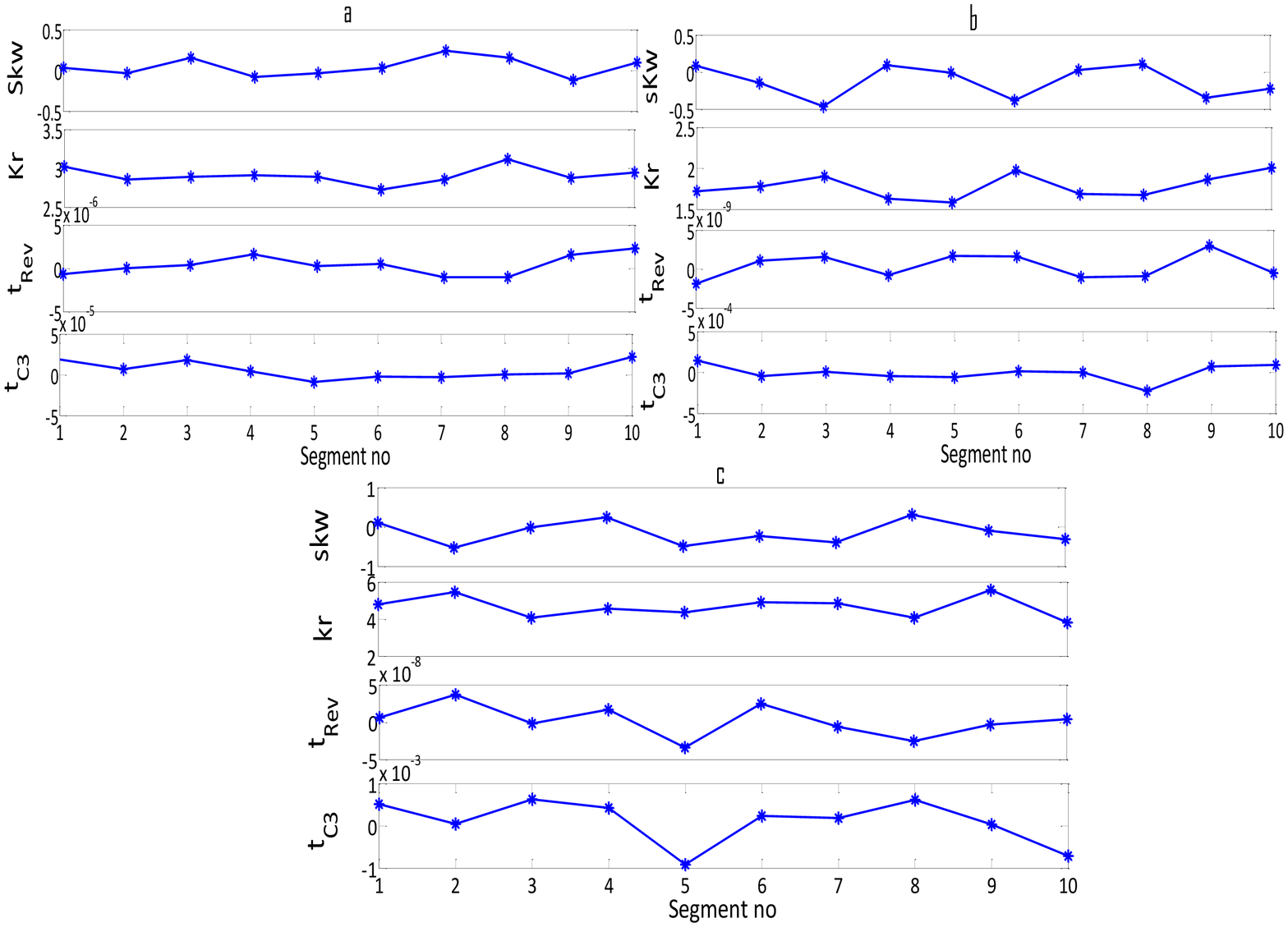}
\caption{plot of kurtosis, skewness, time reversal $t_{rev}$, and third autocovariance $t_{C3}$ at different segments of the time series for a) noise b) sinusoidal c) square forcing with A=2V}
\label{nonstat}
\end{figure}

Here in our work the checking of non-stationarity involves the estimation of certain parameters which in our case is time reversal, third order auto-covariance \cite{dvv1} in different parts of the segments of a time series and if the value of the parameters of nonlinearity measured reveals significantly different value in different segments then the time series can be classified as a non-stationary one. In performing this operation window length of the data containing 1000 points has been chosen to be 10. Shown in Fig. \ref{nonstat} are the estimate of skewness and kurtosis, time reversal $t_{rev}$, and third autocovariance $t^{C3}$  in different segments of a time series with the expressions of the nonlinear parameters like time reversal and third order auto-covariance being given below in equation 4 and \ref{timeauto}.

\begin{eqnarray}
 t^{Rev}(\tau)=<(x_k-x_{k-\tau})^3> \\
 t^{C3}(\tau)=<x_k x_{k-\tau}x_{k-2\tau}>
 \label{timeauto}
 \end{eqnarray}

The presence of nonstationarity permits us to carry out DFA \cite{DFA} for the detection of long range correlations embedded in a nonstationary time series data. Following the approach adapted by \cite{peng} we first integrate the time series  $ y(k)=\sum_{i=1}^k [x(i)-x_{mean}]$ followed by the dividing of the time series into boxes of equal length n. A least square line representing the trend is fitted to the data. A y coordinate of the straight line segment is denoted by $y_n(k)$ in each box. Next we detrend the integrated time series y(k) by subtracting the local trend, $y_n(k)$ in each box. Root mean square fluctuation of this integrated and detrended time series is calculated by

\begin{eqnarray}
F(n)=\sqrt{\frac{1}{N}{\sum_{k=1}^N[y(k)-y_n(k)]^2}}
\label{diagline}
\end{eqnarray}

\begin{figure}
\centering
\includegraphics[width=9cm]{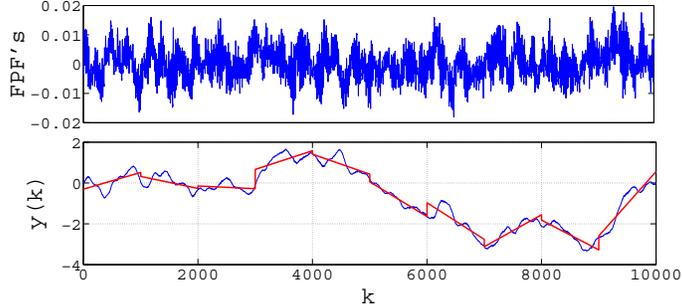}
\caption{Integrated time series superposed on least square fitted trend}
\label{DFA}
\end{figure}

The process is repeated over all time scales to provide a relationship between the average fluctuation as a function of F(n) and the box size n.
In order to produce more accurate result, the largest box size we use is N/10 where N is the total no of points in the FPF's.
A linear relationship on a log log graph indicates the presence of scaling with scaling exponent being $\alpha$ implying that $F(n) \sim n^{\alpha}$. Value of $\alpha$ greater than 0.5 and less than or equal to 1.0 indicates persistent long range correlation and that lying with 0.5 to 1 represent different type of
power law correlation such that large and small values of the time series are likely to alternate. Power law exponent values greater than 1 indicate the existence of perfect correlated dynamics. Altogether the exponent can be viewed as the roughness of the time series, the larger the value of coefficient $\alpha$ the smoother will be the time series.

\section{Results of DFA analysis with crossover phenomena}

\begin{figure}
\centering
\includegraphics[width=13cm, height=8cm]{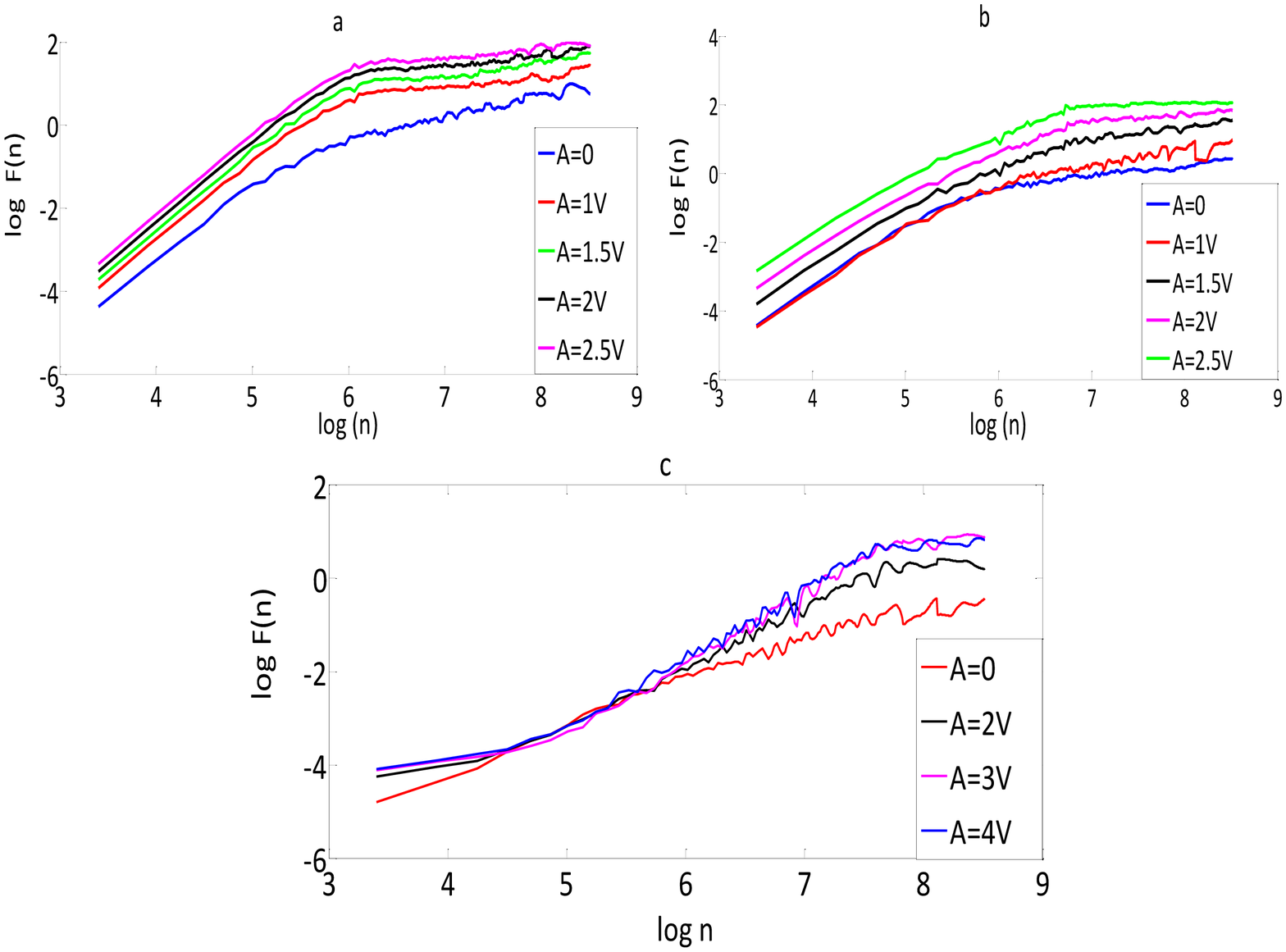}
\caption{Plot of $log F(n)$ vs $log (n)$ for increase in a) sinusoidal forcing b) square forcing  c) sinusoidal forcing on FPF for P=0.004mbar}
\label{DFA1}
\end{figure}

\begin{figure}
\centering
\includegraphics[width=8cm]{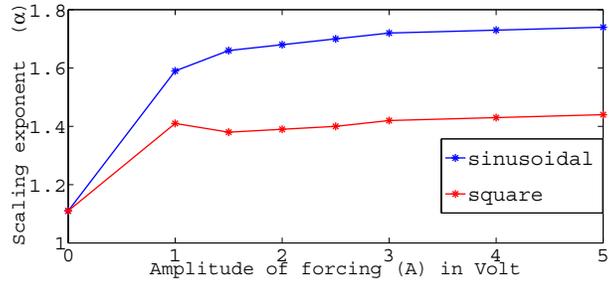}
\caption{Variation of scaling exponent with increase in amplitudes of the sinusoidal as well as square forcing }
\label{DFA_amp}
\end{figure}

\begin{figure}
\centering
\includegraphics[width=8cm]{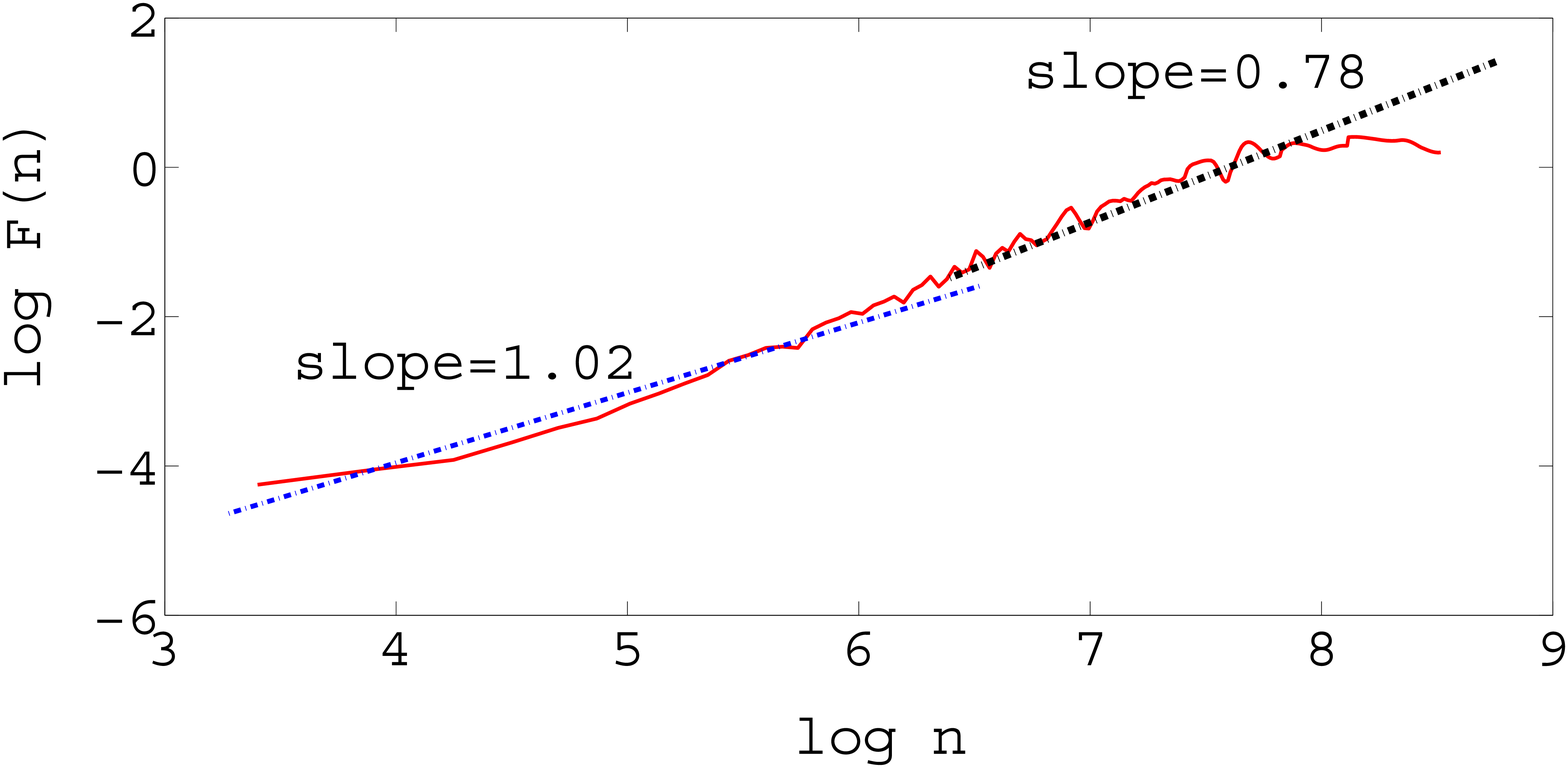}
\caption{Shift in the values of $\alpha$ from low n to high n for sinusoidal forcing applied on fluctuation}
\label{crossover}
\end{figure}

Now we illustrate results in Fig. \ref{DFA1}a and \ref{DFA1}b using the above mentioned technique on the fluctuations acquired with increasing sinusoidal and square forcing amplitudes respectively for P=0.12mbar. We can clearly observe only a main scaling region upto n $\sim$ 403 and beyond that there is a saturation region indicating very small slope. The values of the scaling exponent estimated from the increasing trend of Fig. \ref{DFA1}a, \ref{DFA1}b is clearly portrayed in Fig \ref{DFA_amp}. Irrespective of the nature of forcing the increase in amplitude of the external forcing leads to the increase in scaling exponent$\alpha$ initially from the value of $\sim$ 1.12. Values of $\alpha$ for sinusoidal as well as square forcing amplitudes are seen to increase from A=0 to A=1V followed by the slight increase upto A=3V in Fig. \ref{DFA_amp}. Further increase in A is seen to produce the values of $\alpha$ falling in the saturation region. Shown in the Fig. \ref{DFA1}c is the DFA analysis for increasing sinusoidal forcing amplitude applied on some chaotic FPF's having maximum lyapunov exponent of $\sim$ 0.2 acquired at the condition of very low P=0.004mbar. The slope of the curve allows us to check the scaling exponent for the values of the external forcing with increasing amplitude. Unlike the Fig. \ref{DFA1}a, \ref{DFA1}b here the appearance of two scaling regions is prominently referred to as the crossover type phenomena which indicates shift in the values of $\alpha$ from the lower region of n ($33<n<665$) to the higher region ($665<n<5000$). The two different slopes lying in the two scaling region is displayed separately in Fig. \ref{crossover} and the quantitative values in the two regions are listed in Table I. In the 2nd scaling region we find the values of $\alpha$ to remain within $0.5<\alpha<1$ indicating persistence long range behaviour. Here the $\alpha$ values for the two scaling regions are almost constant at any amplitude of the external forcing.

\begin{figure}
\centering
\includegraphics[width=9cm, height=4cm]{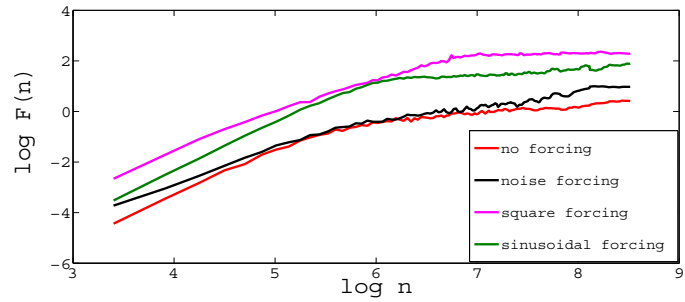}
\caption{Plot of $log F(n)$ vs $log(n)$ for no forcing and different types of forcing i.e noise, sinusoidal, square applied with A=2V}
\label{DFA2}
\end{figure}

\begin{figure}
\centering
\includegraphics[width=7cm, height=5cm]{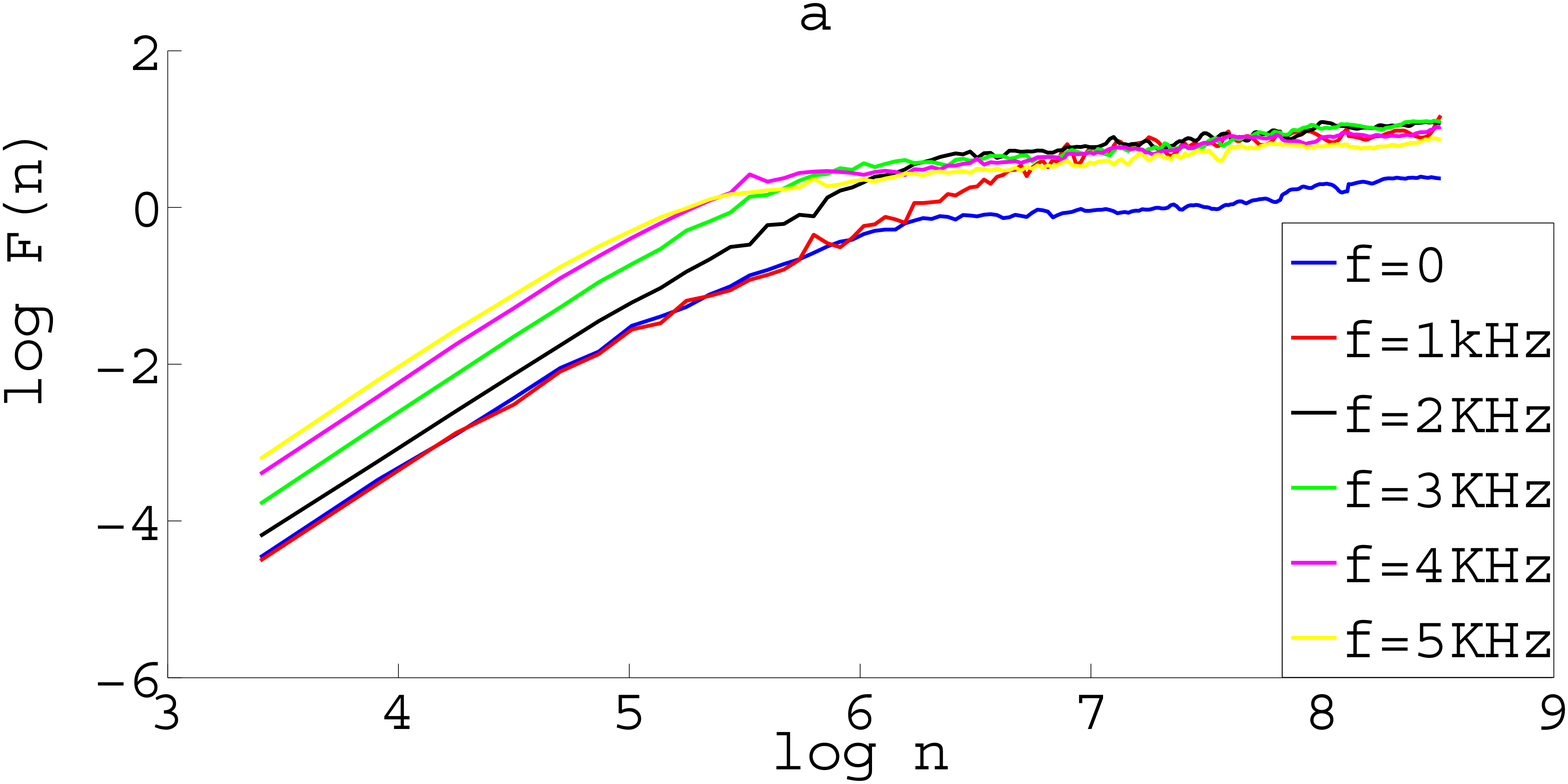}
\includegraphics[width=7cm, height=5cm]{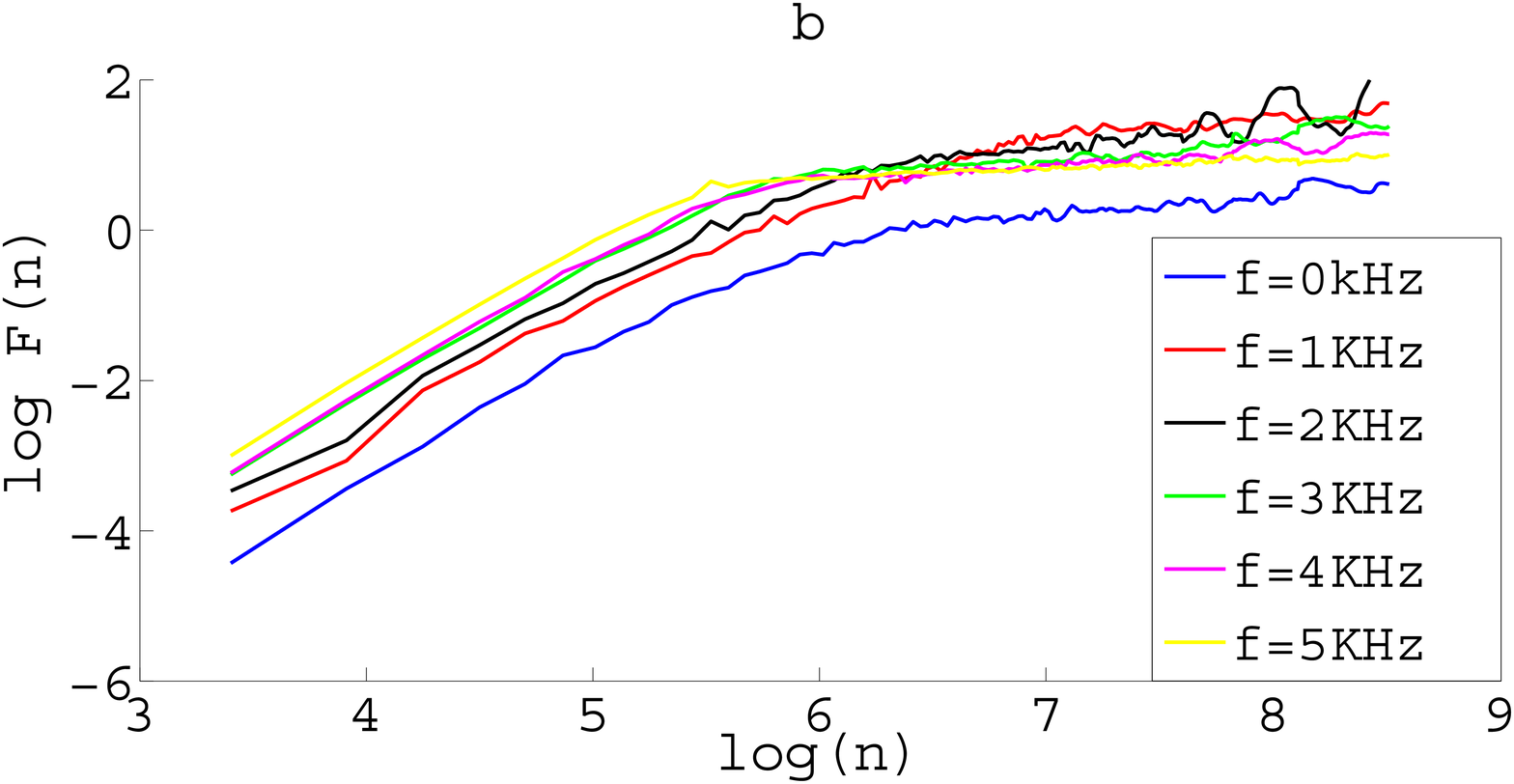}
\caption{Plot of $log F(n)$ vs $log (n)$ for increase in a) sinusoidal forcing (A=2V) and the same plot for increase in b) square forcing with A=2V}
\label{DFA3}
\end{figure}

Fig. \ref{DFA2} represents the DFA analysis for four different time series namely the original data and the rest three generated by applying three different types of forcing i.e noise, sinusoidal, square forcing. The scaling exponent obtained from the slope of F(n) vs n in a double log graph depends paramountly on choosing the value of the box size n. It can be noticed that Fig. \ref{DFA2} exhibit mainly one scaling region or slope for the four curves over the range of n ($33<n<403$) for a fixed A=2Volt. The values of the slope $\alpha$ estimated for the main scaling region are 1.68 (sinusoidal), 1.09 (noise), 1.39 (square) 1.12(no forcing) indicating the correlated dynamics. The smaller values of the scaling exponent for increasing square forcing than the sinusoidal one is also observed from Fig. \ref{DFA_amp} implying the gradual increase in the smoothness of the fluctuations with increase in A of sinusoidal forcing. Now the effect of increasing frequencies of sinusoidal and square forcing keeping A=2V on FPF's acquired under the condition of P=0.12 mbar are displayed in Fig. \ref{DFA3}. The estimated values are listed in table II with the values indicating perfect correlation. Similar to the Fig. \ref{DFA1}a, \ref{DFA1}b and \ref{DFA_amp}(for increasing amplitudes) the initial increase in frequency (sinusoidal, square) lead to increase in scaling exponent values followed by the saturation in its values (Fig. \ref{DFA3}a, Fig. \ref{DFA3}b). In the higher region of the values of n, increase in frequency results in decrease of scaling exponent and the lowest value is observed to be at maximum frequency of 5kHz.

\begin {table}
\caption {Results of the quantitative values of the scaling exponent for forcing applied on FPF's obtained at low P=0.004mbar} \label{tab1}
\begin{center}
\begin{tabular}{| l| l| l|}
\hline
  A (in V)  &  \ \ \ \  slope(region1)  & \ \ \ \  slope(region2)     \\  \hline
  0 &  \ \ \ \  1.01      & \ \ \ \ 0.77     \\  \hline
  1 &  \ \ \ \  1.02      & \ \ \ \ 0.78     \\  \hline
  2 &  \ \ \ \  1.04      & \ \ \ \ 0.77     \\  \hline
  3 &  \ \ \ \  1.06      & \ \ \ \ 0.79     \\  \hline
  4 &  \ \ \ \  1.05      & \ \ \ \ 0.79     \\   \hline

 \end{tabular}
\end{center}
\end {table}

\begin {table}
\caption {Scaling exponent values for increasing frequencies of square and sinusoidal forcing} \label{tab1}
\begin{center}
\begin{tabular}{| l| l| l|}
\hline
  f (in kHz)  &  \ \ \ \  slope(sinusoidal)  & \ \ \ \  slope(square)     \\  \hline
  0 &  \ \ \ \   1.11      &  \ \ \ \ 1.11     \\  \hline
  1 &  \ \ \ \   1.40      &  \ \ \ \ 1.36    \\  \hline
  2 &  \ \ \ \  1.58      &  \ \ \ \ 1.44     \\  \hline
  3 &  \ \ \ \  1.66      &  \ \ \ \ 1.49     \\  \hline
  5 &  \ \ \ \  1.62      &  \ \ \ \ 1.46    \\  \hline

\end{tabular}
\end{center}
\end {table}

\section{\bf Conclusion}

To summarize we carried out DFA analysis to quantify scaling exponent for different amplitudes of sinusoidal, square forcings.
The shift in the values of scaling exponent from one scaling region to the other indicated as crossover phenomena is observed when external forcing is added on fluctuation having lyapunov exponent (LE $\sim$ 0.2) greater than that acquired at P=0.12mbar having LE $\sim$ 0.03. A persistence long range behaviour is revealed in one of these scaling regions with $\alpha$ lying in the range from 0.5<$\alpha$<1. Different nonlinear parameters are estimated in different segments of the time series to detect nonstationarities hence we tried estimating the scaling exponent with the help of DFA analysis in a non-stationary time series. The emergence of scaling region depends on the type of forcing applied. It is worth remembering that in order to carry out DFA analysis a long data set is required for statistically robust result. DFA also helps to study the intrinsic dynamics of a given system by analyzing and correctly interpreting the FPF's which is also executed using RP, RQA. RP is an advanced technique which qualitatively and quantitatively reveals the underlying physics of the system dynamics when the system goes through a transition. The considered recurrence measures exhibit an instantaneous change which was noticed in both RPs and RQA measures through variables like DET, Entropy. The estimated scaling exponents is higher in case of sinusoidal forcings than the square forcing at any value of the amplitude, frequency of applied forcings implying the decreasing roughness of the FPF's for sinusoidal forcing than the square forcing which is reflected also in the Entropy values of RQA analysis. The most striking feature of our observation is how crossovers in the correlation behavior can be detected reliably and determined quantitatively. A crossover is due to the change in the correlation properties of a signal at different time or space scale though it can also be a nonstationarities in the signal. It is the advantage of DFA that it can systematically remove trends embedded in a nonstationary time series and in this way we can gain insight into the scaling behaviour. Correlations in the fluctuations physically imply that they do not decay fast enough and the system possesses long memory. This work highlights the potential of RQA along with the DFA analysis which can be used to explore and develop the dynamical system theory of plasma oscillation of different plasma system like glow discharge, double plasma, dusty plasma. The scaling properties reflecting the inner fractal structures of time series are alternative quantifiers of complexity of dynamical systems. \\

\noindent {\textbf{Acknowledgments:}} The authors would like to acknowledge the director, SINP, for his constant support. The authors are also thankful to technical staff of plasma physics division Dipankar Das and Ashok Ram for technical help during the experiment.

\end{document}